\documentclass[prd,referee,nopacs,nofootinbib, twocolumn]{revtex4}
\usepackage[colorlinks,linkcolor={red},citecolor={blue}]{hyperref}
\usepackage{amsfonts}
\usepackage{amsmath}
\usepackage{eurosym}
\usepackage{amssymb}
\usepackage{graphicx}
\usepackage{color}

\newcommand{\f}[2]{\frac{#1}{#2}}
\def\be{\begin{equation}}
\def\ee{\end{equation}}
\def\bea{\begin{eqnarray}}
\def\eea{\end{eqnarray}}

\begin{document}
	
	\title{Growth of matter density perturbations in 4D Einstein-Gauss-Bonnet gravity}
	\author{Zahra Haghani}
	\email{z.haghani@du.ac.ir}
	\affiliation{School of Physics, Damghan University, Damghan,
		41167-36716, Iran.}
	
	\date{\today}
	
	\begin{abstract}
	Recently, a novel Einstein-Gauss-Bonnet gravity in four dimensions has been introduced [Phys. Rev. Lett. 124 (2020) 081301]. We will investigate cosmological consequences of this model in details. We will consider linear matter density perturbations and also estimate relevant modified gravity parameters. Specially we will concentrate on the growth rate of non-relativistic matter perturbations on top of FRW geometry and show that for a narrow range of the Gauss-Bonnet coupling constant, observational data will be satisfied. In this paper, it is shown that for enough large  values of the coupling constant, the suggested $\sigma_8$ will become greater than $\Lambda$CDM value. 
	\end{abstract}
	
	\maketitle
	\section{Introduction}
	It is well-known that the Einstein-Hilbert term plus cosmological constant is not the unique healthy second order gravitational Lagrangian in higher than four dimensional space-times. The Lovelock theorem \cite{lovelock} states that in $D\geq4$ space-time dimensions, for a generic metric field the unique healthy action with second order field equation can be given by Lovelock-Lanczos action. In four dimensions however, one has left with only the Ricci scalar plus a constant due to the fact that other Lovelock invariants become total derivatives. Generally, in $d+1$ dimensions where $d$ denoted spatial dimensions, one has $(d-1)/2$ (even dimensions) or $d/2$ (odd dimensions) Lovelock invariant terms. From the viewpoint of the field equations, one can prove that the Lovelock tensor in $D$ dimensions has coefficients $(d-n)$, with $n=3,..,d$, making the Lovelock tensor vanishes in lower dimensions \cite{proof}.
	
	Many attempts has been made in the literature to make higher order Lovelock invariants to contribute in four dimensions, including the introduction of extra degrees of freedom, non-minimally coupled to the Lovelock invariants \cite{extradof}, or making non-linear function out of the Lovelock invariants \cite{nonlinear}. 
	
	Recently, a novel way to make the Lovelock tensors  appear in 4 dimensional equations of motion is introduced, which is based on the introduction of non-standard coupling constants. In this approach, we consider the theory in an arbitrary $d+1$ dimensional space-time and modify the coupling constant of the Lovelock invariants to have an extra $(d-3)$ factor. As we have discussed above, the extra prefactor cancels the same factor from the variation of the Lovelock invariant. So, when one takes a limit $d\rightarrow3$, the Lovelock tensor arises in metric field equation.	
	The simplest possibility is to consider the effects of Gauss-Bonnet invariant in four dimensions \cite{PRL}. One can see that the Gauss-Bonnet tensor makes a contribution proportional to $H^4$ in the Friedmann equation. This term would modify the effective cosmological constant of the theory and also shift the Planck mass by a constant while dynamical degrees of freedom of the theory remains 2 \cite{PRL}. Tensor perturbations around FRW geometry is also considered in this context and it is shown that the sound speed and the Hubble friction will be modified by a factor proportional to $H^2/M_P^2$. At late times, $H^2/M_P^2$ is very small and one can expect that the theory can satisfy the recent observations on the gravitational waves produced by merging  two Black holes/Neutron stars \cite{gravitationalwave}.
 Many aspects of the theory has been investigated in the literature, including black holes \cite{blackhole}, and cosmology \cite{cosmo}. Theoretical aspects and generalizations of the idea has also been considered extensively \cite{theory}. The effects of higher order Lovelock terms in four dimensional space-time is considered in \cite{higher} and dynamical system analysis of the theory has also considered. It has been shown that the Friedmann equation is modified by $H^{2n}$ terms where $n$ is the highest order of the Lovelock term in the action.  Cosmological behavior and black hole solutions of this generalized model is also considered \cite{higher}.  
	
	In this paper, we are going to consider the effects of this new 4D Gauss-Bonnet term on the growth rate of the non-relativistic matter density perturbations. We have assumed that the matter content of the universe can be described by a perfect fluid with barotropic equation of state $p=\omega\rho$. We will show that the Gauss-Bonnet term modifies the behavior of Hubble parameter and the acceleration rate of the universe at early times. Matter density perturbations around the FRW geometry show that the Gauss-Bonnet term will affect the growth rate at early times and also modify the $f\sigma_8$ value. 
	
	The paper is organized as follows: In the next section, we will give a brief review of the theory. In section \ref{cosmology}, the background cosmological evolution of the theory in the presence of dust and radiations is considered. In section \ref{pert} we will study the pressureless matter density perturbations of the theory in subhorizon limit and investigate the effects of the 4D Gauss-Bonnet term in the growth rate of the matter perturbations. Section \ref{conclusion} will be devoted to conclusions and final remarks.
\section{The Model}\label{model}
The 4D Einstein-Gauss-Bonnet (EGB) action in (d+1) dimensional space time is
\begin{align}\label{act}
S=\int& d^{d+1}x\sqrt{-g}\bigg(\kappa^2(R-2\Lambda)+\mathcal{L}_m+\f{\alpha}{d-3}\mathcal{G}\bigg),\
\end{align}
where $\Lambda$ is the cosmological constant, $\mathcal{L}_m$ is the Lagrangian of the matter field, $d$ is the dimension of the constant-time hypersurfaces and $\mathcal{G}$ is the Gauss-Bonnet Lagrangian defined as
\begin{align}\label{GB}
\mathcal{G}=R^{\mu\nu\alpha\beta}R_{\mu\nu\alpha\beta}-4R^{\mu\nu}R_{\mu\nu}+R^2.
\end{align}
As was discussed in the Introduction, the coupling constant of the Gauss-Bonnet term is written in such a way that it cancels the $d-3$ factor in the Gauss-Bonnet tensor.
One can obtain the field equations of the action \eqref{act} as
\begin{align}\label{fe}
2\kappa^2 &\left(G_{\mu\nu}+\Lambda g_{\mu\nu}\right)+\f{\alpha}{d-3}\big(4R R_{\mu\nu}-8R_{\mu\alpha}R^{\alpha}_\nu\nonumber\\&-8R_{\mu\alpha\nu\beta}R^{\alpha\beta}+4R_{\mu\alpha\beta\sigma}R_{\nu}^{~\alpha\beta\sigma}
-g_{\mu\nu}\mathcal{G}\big)
=T_{\mu\nu},
\end{align}
where the energy-momentum tensor $T_{\mu\nu}$  of the matter field is defined as
\begin{align}
T_{\mu\nu}=-\frac{2}{\sqrt{-g}}\frac{\delta \left(\sqrt{-g}\mathcal{L}_m\right)}{\delta g^{\mu\nu}}.
\end{align}
The expression in the second parenthesis of equation \eqref{fe} is the Gauss-Bonnet tensor with the property that it contains at most second order derivatives. As a result the above theory is free from Ostrogradski instability. 

One can easily verify that the energy-momentum tensor of the matter field is conserved
\begin{align}\label{cons}
\nabla_\mu T^{\mu\nu}=0,
\end{align}
due to the Bianchi identity applied to the Gauss-Bonnet tensor. This is however trivial since we do not introduce non-minimal matter geometry couplings in the action.
\section{Cosmology}\label{cosmology}
In this section, we will investigate the cosmological implications of the 4D EGB theory by adopting the homogeneous flat  FRW ansatz for the geometry of the Universe, given by
\begin{align}
ds^2=a(t)^2\eta_{\mu\nu}dx^\mu dx^\nu,
\end{align}
where $a(t)$ is the scale factor, the parameter $t$  stands for the conformal time and $\eta_{\mu\nu}$ is the Minkowski metric. We assume that the Universe is filled with perfect fluid which is characterized  by energy density $\rho$ and thermodynamic pressure $p$, with the energy-momentum tensor given by
\begin{align}
T^{\mu\nu}=(\rho+p)u^\mu u^\nu +p g^{\mu\nu}.
\end{align}
We will also assume that the equation of state for the matter sources is the barotropic equation of state with  $p=\omega \rho$, where $\omega$ is a constant.

 With these assumptions, the Friedmann and Raychadhuri equations for 4D EGB theory reduce to
\begin{align}\label{frid}
2\kappa ^2 \left(3 \mathcal{H}^2- a^2 \Lambda \right)+\frac{6 \alpha  \mathcal{H}^4}{a^2}=a^2 \rho,
\end{align}
and
\begin{align}\label{ray}
2 \kappa ^2 \left(2\dot{\mathcal{H}}-a^2 \Lambda +\mathcal{H}^2\right)-\frac{2 \alpha}{a^2} \mathcal{H}^2
	\left(\mathcal{H}^2-4 \dot{\mathcal{H}}\right)=-a^2 p,
\end{align}
where $\mathcal{H}=\f{\dot{a}}{a}$ is the Hubble parameter and dot denotes derivative with respect to the conformal time $t$. The conservation equation of the matter field \eqref{cons} can be written as
\begin{align}\label{cons1}
\dot{\rho} +3 \mathcal{H} (p+\rho )=0.
\end{align}
Now, suppose that the Universe is filled with non relativistic matter with  $\omega_m=0$ and the relativistic matter with $\omega_r=1/3$ such that
\begin{align}\label{dens}
\rho=\rho_m+\rho_r, \quad p=p_r=\f13 \rho_r.
\end{align}
Let us define the following set of dimensionless parameters
\begin{align}\label{dimless}
&\tau=H_0 t,\quad \mathcal{H}=H_0 h, \quad \beta=\f{H_0^2}{\kappa^2}\alpha, \nonumber\\&
\bar{\rho}_{\Lambda}=\f{1}{3H_0^2}\Lambda,\quad  \bar{\rho}_i=\f{1}{6 \kappa^2 H_0^2}\rho_i,~~i=r,m.
\end{align}
where $H_0$ is the current Hubble parameter. 
We suppose that the radiation and non-relativistic matter are conserved separately, so  from equation \eqref{cons} we can obtain the behavior of $\bar{\rho}_i$'s in terms of the scale factor as 
\begin{align}
\bar{\rho}_r=\f{\Omega_{r0}}{a^4}, \quad \bar\rho_m=\f{\Omega_{m0}}{a^3}, 
\end{align}
where the constants $\Omega_{r0}$ and $\Omega_{m0}$ are the present time density parameters of radiation and dust, respectively. The vales of these parameters from the Plank data \cite{plank} are $\Omega_{m0}=0.305$ and $\Omega_{r0}=0.531\times 10^{-4}$.

To compare the cosmological behavior of the model with cosmological observations we use the redshift parameter $z$ instead of the conformal time defined as
\begin{align}\label{zz}
1+z=\f1a,
\end{align}
where we have used the normalized scale factor by taking $a(0)=1$. Using the relations \eqref{dimless} and \eqref{zz}, one can rewrite the field equations \eqref{frid} and \eqref{ray}   as
\begin{align}\label{fridz}
 (z+1)^2\left(\beta  h^4- \Omega
 _{\text{r0}}\right)+h^2-(z+1) \Omega _{\text{m0}}-\frac{\bar{\rho}_\Lambda}{(z+1)^2}=0,
\end{align}
and
\begin{align}\label{rayz}
h^2&-2 h (z+1) h'-\beta  h^3 (z+1)^2 \left(4 (z+1) h'+h\right)\nonumber\\&-\frac{3
	\bar{\rho}_\Lambda }{(z+1)^2}=0,
\end{align}
where prime denotes derivative with respect to the redshift $z$.

\begin{figure}[h]
	\centering
	\includegraphics[scale=0.45]{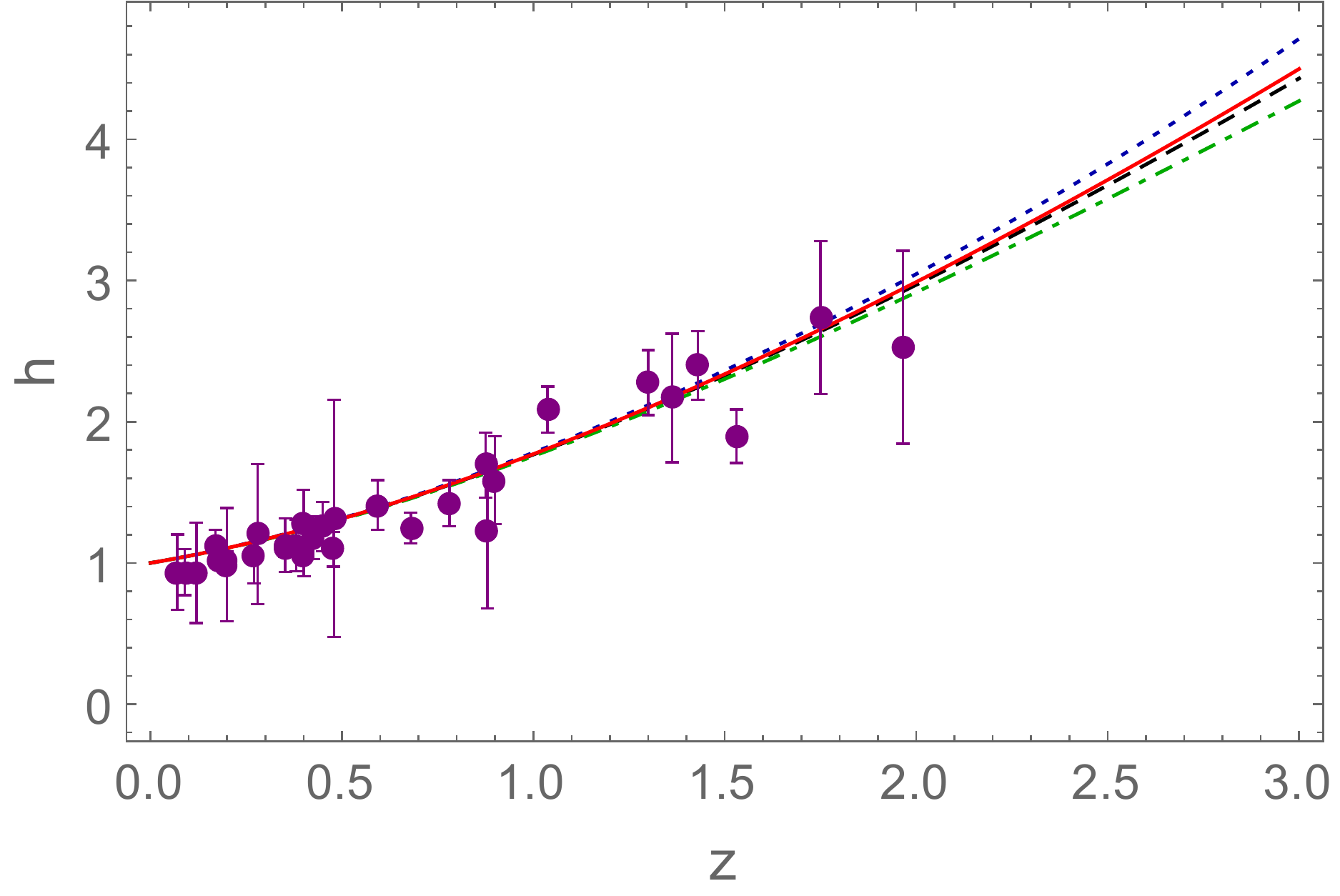}
	\caption{Variation of the Hubble parameter $h$ as a function of redshift $z$ for  different values of the parameter $\beta$, $\beta=0.0015$ (dashed curve), $\beta=-0.004$ (dotted curve), $\beta=0.006$ (dot-dashed curve) and $\Lambda$CDM (solid curve). The observational data are shown by error bars.}\label{h}
\end{figure}

\begin{figure}[h]
	\centering
	\includegraphics[scale=0.45]{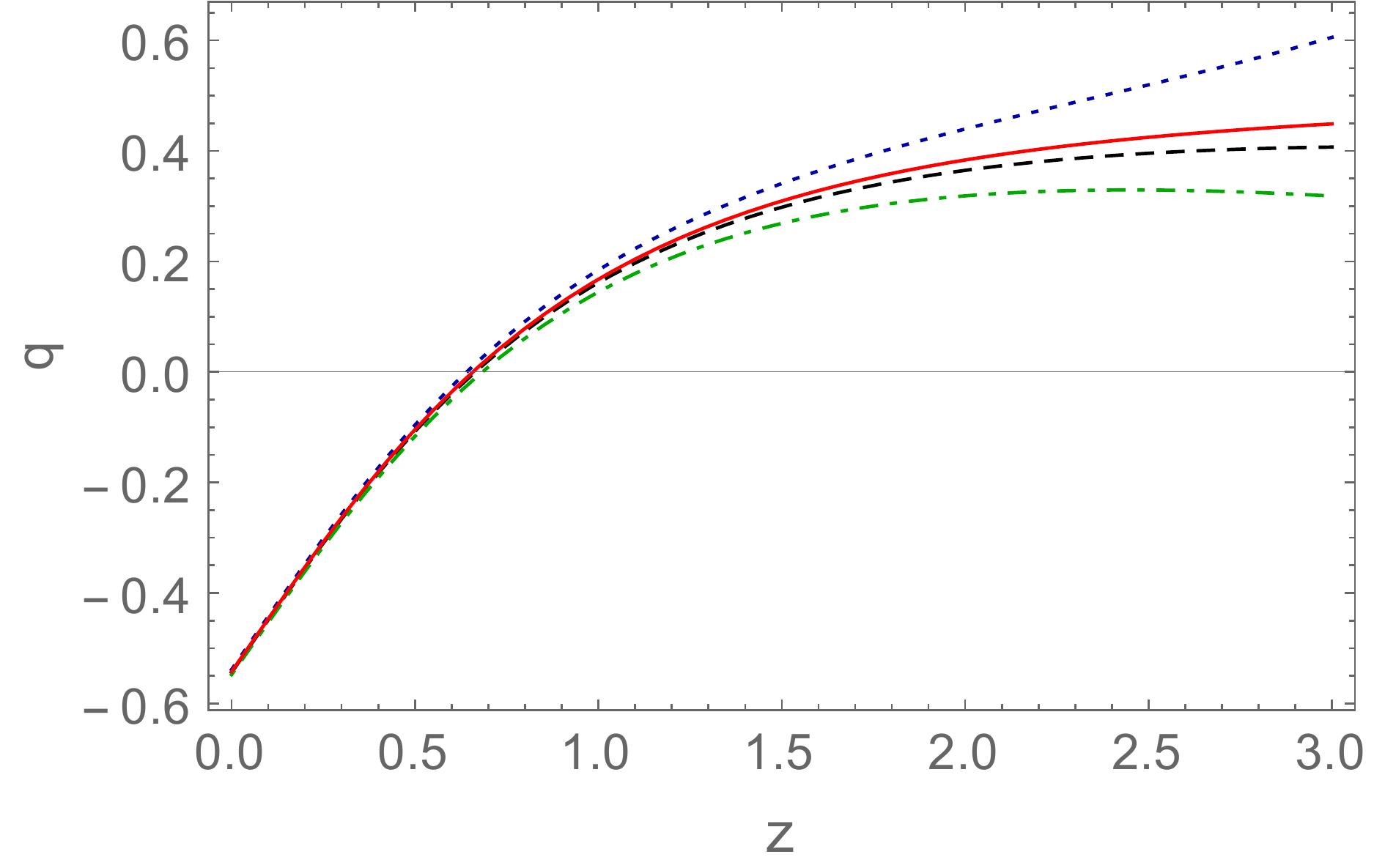}
	\caption{Variation of  deceleration parameter $q$ as a function of redshift $z$ for  different values of the parameter $\beta$,  $\beta=0.0015$ (dashed curve), $\beta=-0.004$ (dotted curve), $\beta=0.006$ (dot-dashed curve)  and $\Lambda$CDM (solid$  $ curve).}\label{q}
\end{figure}

To investigate the cosmological evolution of the Universe,  considering the evolution of the Hubble parameter $h$ and of the deceleration parameter $q$ are necessary. The deceleration parameter  determines that the expansion of the Universe is whether accelerating or decelerating. In terms of the redshift $z$ this parameter can be obtained as
\begin{align}
q=\left(1+z\right)\f{d \ln h}{dz}.
\end{align}
To investigate the evolution of the cosmological parameters $h$ and $q$ in the 4D EGB theory, we consider the field equations \eqref{fridz} and \eqref{rayz}. There are two free dimensionless parameters $\bar{\rho}_\Lambda$ and $\beta$ in this model. By the use of Friedmann equation at the present time $z=0$, and adopting the $h(0)=1$, one can easily obtain the value of the parameter $\bar\rho_\Lambda$ in terms of $\beta$. To find the best-fit parameter value, we will use the likelihood analysis in the next section. The observational data  points for Hubble parameter are in \cite{hobs}. These data are obtained by measuring the age difference between two ensembles of passively evolving galaxies at different redshifts, one could determine the derivative of redshift with respect to cosmic time. This leads to determine the Hubble parameter. The likelihood analysis in the next section will show that  best-fit value of parameter $\beta$ is $\beta=0.0015$. The  $1\sigma$ and   $2\sigma$ confidence interval  for this parameter is shown in Table \ref{T1}.
The variation of the Hubble Parameter $h$  for three different values of $\beta$ in  $2\sigma$  region  with respect to the redshift $z$ are depicted in FIG. \ref{h}.
 One can see from the figure that the late time behavior of the Hubble parameter is independent on the value of the model parameter $\beta$ and in this era the evolution of this parameter is well matched with  the standard $\Lambda$CDM model and observational data. 
In FIG. \ref{q}, the variation of the deceleration parameter as a function of redshift for different values of the parameter $\beta$ is shown. The solid curve shows the behavior of the deceleration parameter in  the $\Lambda$CDM model. This figure shows that the deceleration parameter has a transition from a decelerated to an accelerated phase. The transition takes place at the same time of the transition in the  $\Lambda$CDM model. At late times, the behavior of the deceleration parameter is independent of the value of $\beta$ and the evolution of $q$ in the $\Lambda$CDM model is reproduced. At early times the values of $\beta$ affects the evolution of the deceleration parameter. It should be noted that larger values of the model parameter $\beta$ can even make the universe to accelerate at high redshifts. 


In the next section, we will consider the matter density perturbations of the model around FRW background and compare them with the observational data.
\section{Matter density perturbations}\label{pert}
In this section, we will consider the scalar perturbations of the field equations \eqref{fe} and \eqref{cons} in the Newtonian gauge to obtain the matter density perturbation of the 4D EGB model. 
The  perturbed conformal FRW metric in Newtonian gauge can be written as
\begin{align}
ds^2=a^2(t)\Big[-(1+2\varphi)dt^2+(1-2\psi)d\vec{x}^2\Big],
\end{align}
where $\varphi$ and $\psi$ are the metric perturbations. 
The perturbed energy momentum tensor of ordinary matter is defined as
\begin{align}
&\delta T^0_0=-\delta\rho\equiv-\rho\,\delta,\quad \delta T^0_i=(1+\omega)\rho\,\partial_i v,\quad\nonumber\\& \delta T^i_j=\delta^i_jc_s^2\rho\,\delta,
\end{align}
where, $\rho$ is the background density, $\delta$ is the matter density contrast defined as $\delta=\delta\rho/\rho$ and $v$ is the scalar mode of the velocity perturbation.
We assume that the perturbations of the pressure is given by  $\delta p/\delta\rho=c_s^2$, where $c_s$ is the adiabatic sound speed of the fluid. The equation of state at the background level is $\omega=p/\rho$. 

In the following, we will restrict ourselves to the matter density perturbations of the pressureless matter source. As a result we expect that the equation of state parameter and the sound speed are zero, so we set $\omega=0=c_s$. With this assumption we use the notation $\delta_m\equiv \delta$ for matter density perturbation in the matter dominated epoch. Also for the ease of calculations, in the following we will Fourier transform the perturbed equation.

\begin{figure}[h] 
 	\centering
 	\includegraphics[width=8.4cm]{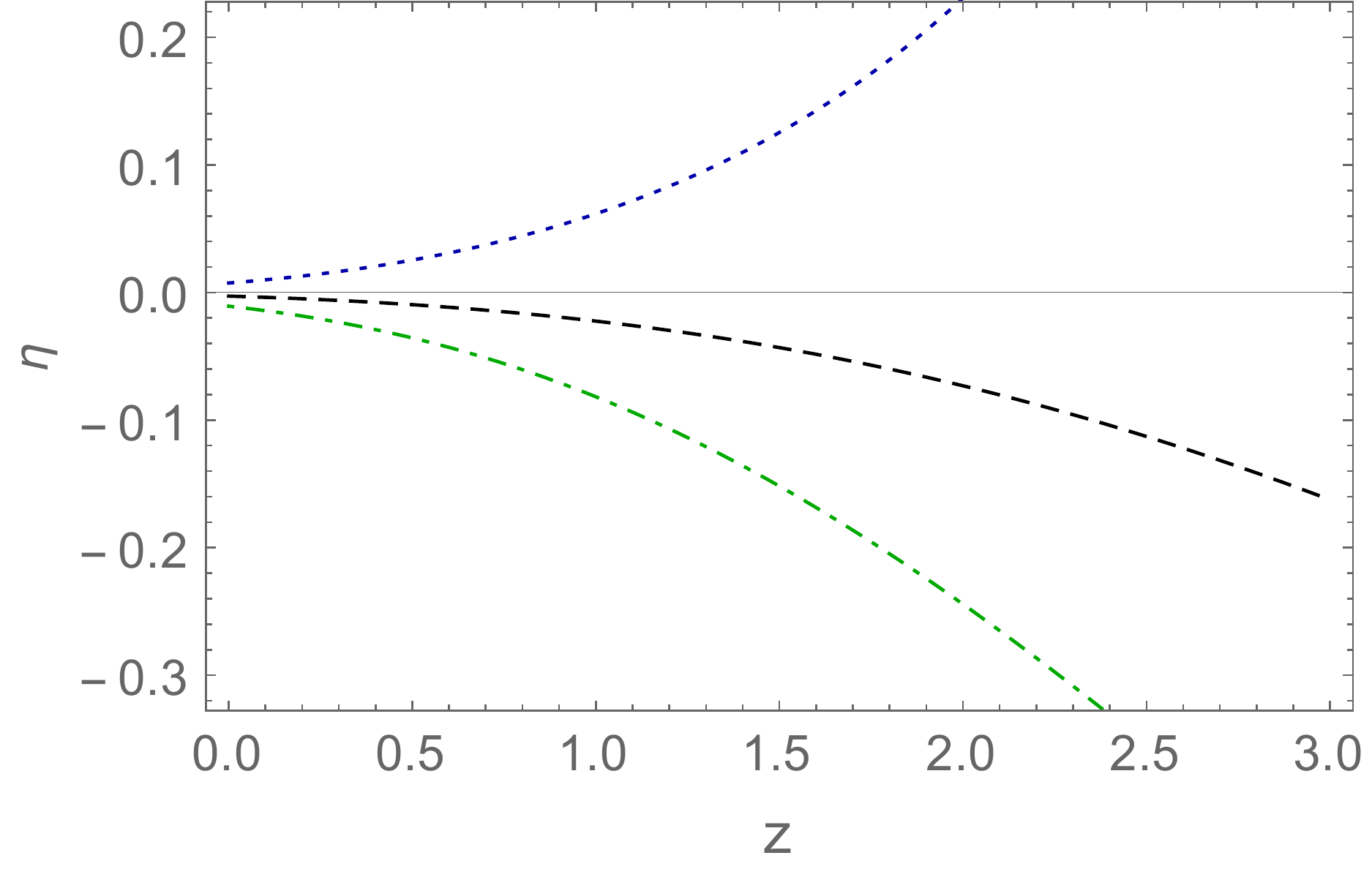}
 	\caption{Variation of the anisotropic stress parameter $\eta$ as a function of redshift $z$ for  different values of the parameter$\beta$, $\beta=0.0015$ (dashed curve), $\beta=-0.004$ (dotted curve), $\beta=0.006$ (dot-dashed curve).}\label{eta}
 \end{figure}

Because the energy-momentum tensor in this model is conserved, the perturbation of the conservation equation \eqref{cons} is identical to the one in the Einstein general relativity. The perturbed temporal and spatial components of equation \eqref{cons} can be written as
 \begin{align}\label{cons0}
 \theta=3\dot{\psi}-\dot{\delta}_m,
 \end{align}
 and
 \begin{align}\label{consi}
 \dot{\theta}+\mathcal{H}\theta-k^2\varphi=0,
 \end{align}
 where $\theta=\nabla_i \nabla^i v$ is the velocity divergence.  
 The $(00)$ component of the metric field equation is
 \begin{align}\label{00} 
 a^4 \rho\,  \delta _m+4 \mathcal{A}\big(k^2 \psi +3 \mathcal{H}^2 \varphi +3 \mathcal{H}
 \dot{\psi} \big)=0,
 \end{align}
 where we have defined
 \begin{align}
 &\mathcal{A}=2 \alpha  \mathcal{H}^2+a^2 \kappa ^2,\nonumber\\&\mathcal{B}=2 \alpha  \mathcal{H}^2-a^2 \kappa ^2.
 \end{align}
The spatial off-diagonal component  of equation \eqref{fe} leads to the relation
 \begin{align}\label{ij}
 \mathcal{A}\varphi+\big(\mathcal{B}-4 \alpha   \dot{\mathcal{H}}\big)\psi=0.
 \end{align}
Using above equation, one can obtain the anisotropic stress parameter as
 \begin{align}
 \eta=\f{\varphi-\psi}{\psi}=\f{4 \alpha}{\mathcal{A}}\left(\dot{\mathcal{H}}-\mathcal{H}^2\right).
 \end{align}
 It should be noted that the anisotropic stress parameter becomes zero in the case of vanishing $\alpha$. The evolution of anisotropic stress parameter as a function of redshift is depicted in FIG. \ref{eta} for three different values of $\beta$.
 It can be seen from the figure that at early times, the deviations of the anisotropic stress parameter
 from zero becomes large. Since the observational data for this parameter are still weak \cite{17}, it can not be used to restrict modifications of the gravitational field.

 The $(ii)$ component of the metric field equation is
 \begin{align}\label{ii}
  &2 \mathcal{H} \mathcal{A}\dot{\varphi}+\Big(4 \dot{\mathcal{H}} \left(2 \mathcal{A}-a^2 \kappa ^2\right)-k^2 \mathcal{A}-2 \mathcal{H}^2 \mathcal{B}\Big)\varphi
 \nonumber\\&+2 \mathcal{A}\ddot{\psi}+4 \mathcal{H}   \big(a^2 \kappa ^2+2 \alpha  \dot{\mathcal{H}}\big)\dot{\psi}+k^2  \big(4 \alpha  \dot{\mathcal{H}}-\mathcal{B}\big)\psi =0,
 \end{align}
 
 We are interested in the evolution of the matter density perturbation $\delta_m$. In the following we will restrict our considerations to the sub-horizon scales in which the Hubble radius is much greater than the physical wavelength. 
\subsection{Subhorizon limit}
\begin{figure}[h]
	\centering
	\includegraphics[width=8.4cm]{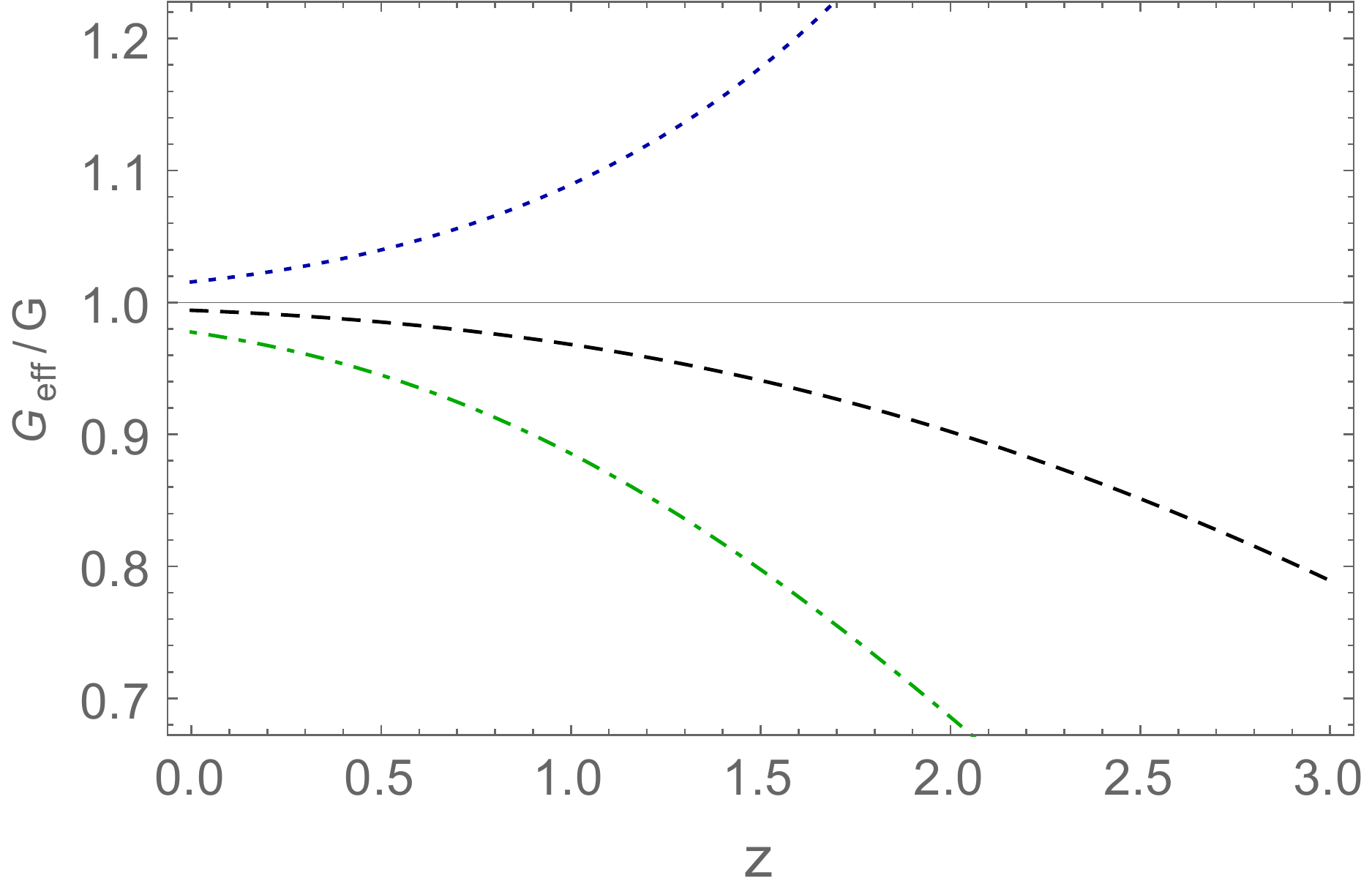}
	\caption{Variation of the  $G_{eff}/G$ as a function of redshift $z$ for  different values of the parameter $\beta$, $\beta=0.0015$ (dashed curve), $\beta=-0.004$ (dotted curve), $\beta=0.006$ (dot-dashed curve).}\label{geff}
\end{figure}

In the subhorizon limit, where $\mathcal{H}\ll k/2\pi a$, equation \eqref{00} takes the form
\begin{align}
a^4 \rho\,  \delta _m+4 \mathcal{A}k^2 \psi=0.
\end{align}
Using equation \eqref{ij} one can obtain the generalized Poisson equation as
\begin{align}
 \rho\,  \delta _m+4 \f{k^2}{a^2} \f{\mathcal{A}^2}{a^2(4 \alpha\dot{\mathcal{H}}-\mathcal{B})}\varphi=0.
\end{align}
This relation can be written in Fourier space as
\begin{align}
\varphi=-4\pi G_{eff}\f{a^2}{k^2}\rho\,  \delta _m,
\end{align}
where we have defined the effective gravitational constant as
\begin{align}
	\f{G_{eff}}{G}=\f{\kappa^2 a^2(4 \alpha\dot{\mathcal{H}}-\mathcal{B})}{\mathcal{A}^2}.
\end{align}
The deviation of $G_{eff}$ from the Newtonian gravitational constant $G$ in terms of redshift is shown in  FIG. \ref{geff}. One can see that at $z> 0.5$, the effective gravitational constant differs significantly with the Newtonian value. 
It should be noted that for both quantities $\eta$ and $G_{eff}$, one can see that at late times their values become approximately equal to the standard $\Lambda$CDM value. As a result, one expects that the qualitative behavior of our model will be identical to the $\Lambda$CDM model at these times.

By using the dimensionless parameters, equation \eqref{00} in the subhorizon limit in terms of redshift yields	
\begin{align}\label{phidel}
\frac{2k^2}{3 H_0^2}&\left(2 \beta  h^2 (z+1)^2+1\right)^2 \varphi = (z+1) \Omega _{m0}\nonumber\\ &\times \left(2 \beta  h^2 (z+1)^2 + 4 \beta  h (z+1)^3 h'-1 \right)\delta_m.
\end{align} 
By substituting \eqref{ij} into \eqref{cons0} one can easily obtain $\theta$ in terms of the matter density contrast  $\delta_m$ and the Newtonian potential $\varphi$. Using the normalized parameter 
\begin{align}
D(z)=\f{\delta_m(z)}{\delta_m(0)},
\end{align}
one can obtain a differential equation for normalized matter density contrast as
\begin{align}\label{maindel}
D ''&+\frac{h'}{h}D' \nonumber\\&+\frac{3  \Omega _{\text{m0}} \left(2 \beta  h^2 (z+1)^2+4 \beta  h (z+1)^3 h'-1\right)}{2
	(z+1) \left(2 \beta  h^3 (z+1)^2+h\right)^2}D =0.
\end{align}

To solve the above equation we use the
 initial condition as 
\begin{align}
\f{d D}{d\ln a}|_{z_\star}=\gamma D|_{z_\star},
\end{align}
where $\gamma$ is an arbitrary constant. It should be noted that the same initial condition with $\gamma=1$ is used in the $\Lambda$CDM model at the deep matter dominated phase at redshift $z=z_\star$. For modified gravities the generalized initial condition parameter $\gamma$ shows the deviation from the $\Lambda$CDM initial condition. In the following we choose  $z_\star=7.1$.

The growth rate of matter perturbation is defined as
\begin{align}
f=\f{d \ln\delta_m}{d\ln a}=-(1+z)\f{D'}{D}.
\end{align}
Equation \eqref{maindel} can be written in terms of $f$ as
\begin{align}
f'&+\left(\f{h'}{h}-\f{1+f}{1+z}\right)f\nonumber\\&-\frac{3  \Omega _{\text{m0}} \left(2 \beta  h^2 (z+1)^2+4 \beta  h (z+1)^3 h'-1\right)}{2
	 \left(2 \beta  h^3 (z+1)^2+h\right)^2}=0,
\end{align}
with the initial condition $f(z_\star)=\gamma$.

To consider the compatibility of the model with observations we use the available data for $f\sigma_8$ \cite{fsig8}. This parameter is defined as
\begin{align}
	f \sigma_8 \equiv \sigma_8(z)\f{D'}{D},
\end{align}
where $\sigma_8(z)=\sigma_8^0\, D(z)$. The constant $\sigma_8^0$ is a model dependent parameter. So to determine this parameter from observations we should at first specify the underlying model. The observational data from weak lensing \cite{wl}, CMB power spectrum \cite{pcmb} and abundance of clusters \cite{abclus} can determine the value of  $\sigma_8^0$. In the following we will find the value of  $\sigma_8^0$ for the 4D EGB model using the likelihood analysis.

To solve the equation \eqref{maindel} we need two initial conditions. Consequently we have two free parameters $\gamma$ and $\sigma_8^0$. There is also the model parameter $\beta$ which we can determine it to obtain the best-fit with observational data. This parameter also appears in Friedmann equations \eqref{fridz} and \eqref{rayz}.  We will determine the values of $\sigma_8^0$, $\gamma$ and $\beta$ by the likelihood analysis \cite{likli}. We have used table II of the paper \cite{fsig8} for the observational data points  $f\sigma_8$ and  the observational data  points for Hubble parameter in \cite{hobs}. In the case of independent data sets, the likelihood functions for the observational data for $H$ and $f\sigma_8$ can be defined as
\begin{align}
\mathcal{L}=\mathcal{L}_{0}e^{-\chi^2/2},
\end{align}
where $\mathcal{L}_0$ is the normalization constant and the quantity $\chi^2$ in our case  is given by
\begin{align}
\chi^2&=\chi_H^2+\chi_{f\sigma_8}^2\nonumber\\&=\sum_i\left(\f{H_{i,obs}-H_{i,teor}}{\sigma_i}\right)^2\nonumber\\&~~~~+\sum_j\left(\f{f\sigma_{8j,obs}-f\sigma_{8j,teor}}{\sigma_j}\right)^2,
\end{align}
where  $H_{i,teor}$ and  $f\sigma_{8j,teor}$ are the theoretical values  for the observables $H_{i,obs}$ and $f\sigma_{8j,obs}$ respectively and $\sigma_i$ is the error associated with the $i$th data. 
The quantity $H_{i,teor}$ depends on $\beta$ and the values of $f\sigma_{8j,teor}$ is in terms of the constants $\beta$, $\gamma$ and $\sigma_8^0$.
By maximizing the likelihood function with respect to these parameters, one can find the best-fit values of  parameters which is presented in Table \ref{T1} with $1\sigma$ and $2\sigma$ confidence levels.	
\begin{table}[h!]
	\begin{center}
		\begin{tabular}{|c||c||c||c|} 
			\hline
			$~~~~~~~~~~$& ~~best-fit~~&$~~~~1\sigma~~~~$&$~~~~2\sigma~~~~$ \\
			\hline
			\hline
			$~~~\beta~~~$ & 0.0015&$\pm$0.0035 & $\pm$0.0096   \\
			\hline
			$\sigma_8^0$ & 0.7295 &$\pm 0.0176$  & $\pm$0.0069 \\
			\hline
			$\gamma$ &3.8652& $\pm$0.1734 & $\pm$0.3399 \\
			\hline
		\end{tabular}
		\caption{The best-fit values of parameters $\beta$, $\sigma_8^0$ and $\gamma$ with $1\sigma$ and $2\sigma$ confidence levels.}\label{T1}
		\label{tab:table1}
	\end{center}
\end{table}
At the best-fit point we have $\chi_{f\sigma_8}^2/dof=0.437$.

\begin{figure}[h]
	\centering
	\includegraphics[width=8.4cm]{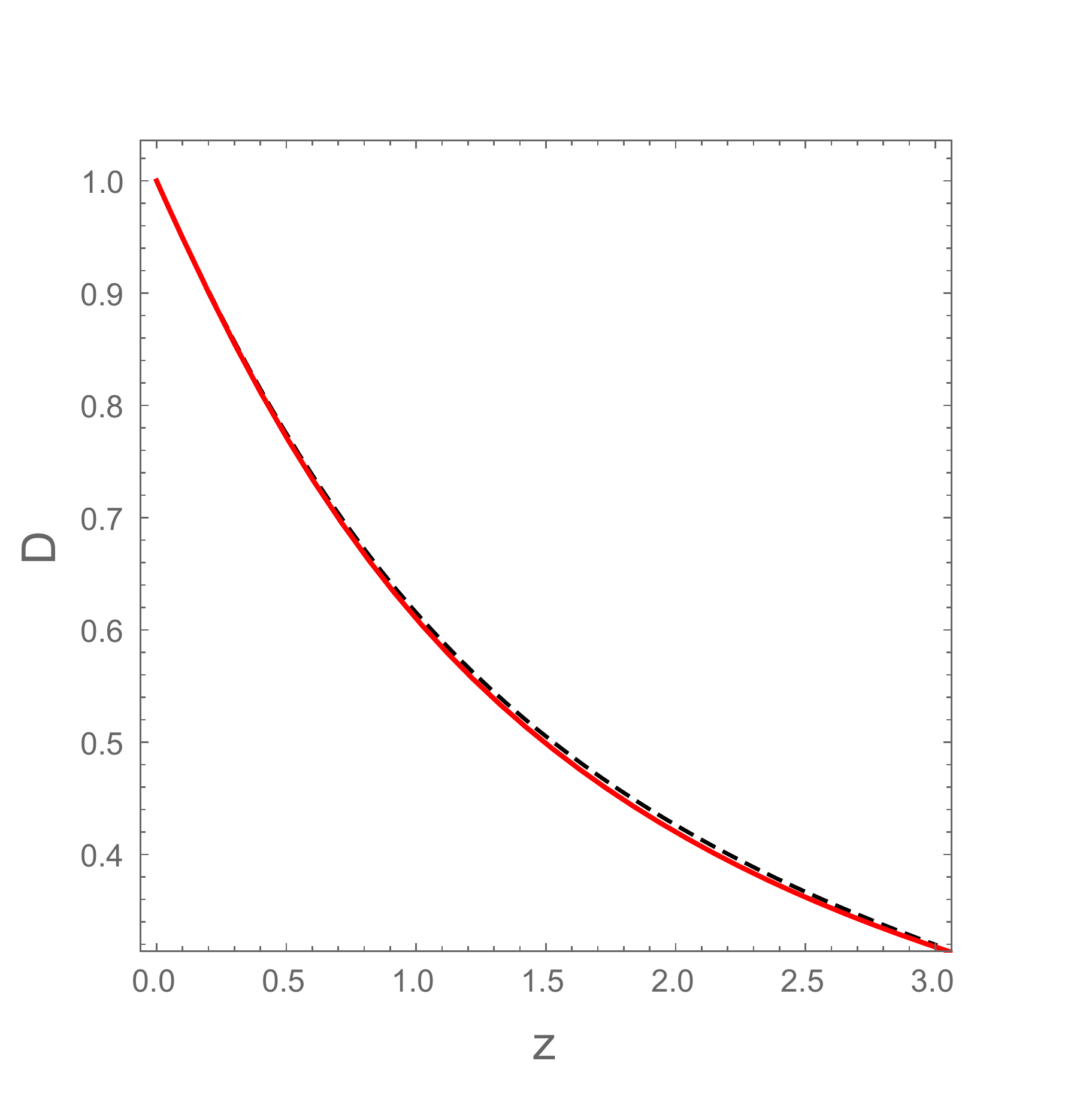}
	\caption{Variation of the normalized matter density perturbation $D$  as a function of redshift $z$ for the best-fit values of parameters in Table \ref{T1} (dashed curve)  and $\Lambda$CDM (solid curve).}\label{delta}
\end{figure}

In FIG. \ref{delta} we have plotted the evolution of the matter density contrast for the best-fit value of parameters in terms of redshift $z$. The red solid curve shows the variation of normalized matter density perturbation for $\Lambda$CDM model. This figure shows the well-matched of the matter density contrast  in this model with $\Lambda$CDM.
In FIG. \ref{fs8}, we have plotted $f\sigma_8$  for 4D EGB model with the best-fit value of  parameters presented in the Table \ref{T1}. In this figure the red solid line corresponds to the $\Lambda$CDM model and  the error bars show the observational data for $f\sigma_8$ \cite{fsig8}. 
\textsc{\begin{figure}[h!]
		\centering
		\includegraphics[width=8.5cm]{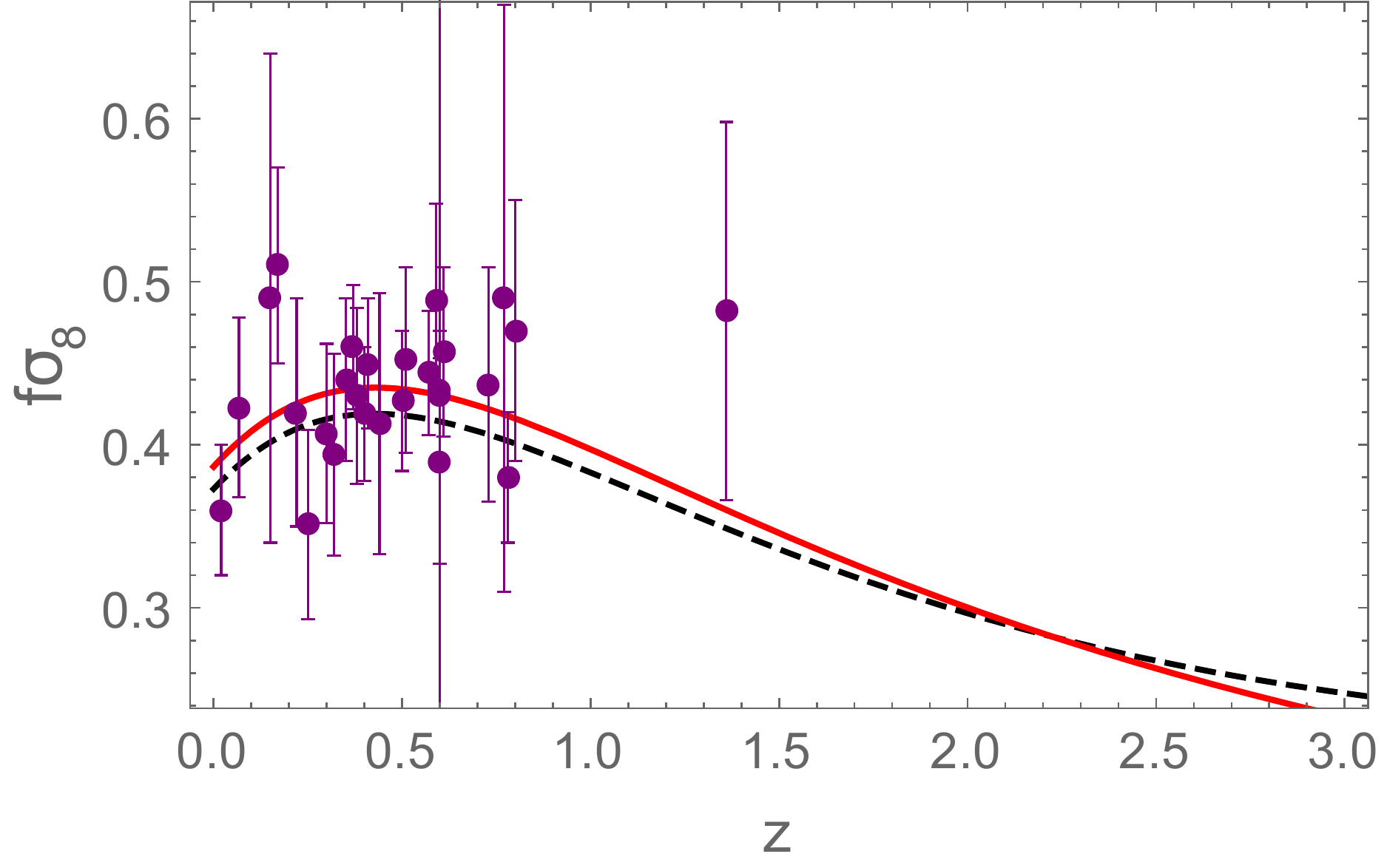}
		\caption{Variation of the  $f\sigma_8$  as a function of redshift $z$ for  the best-fit values of parameters in Table \ref{T1} (dashed curve) and $\Lambda$CDM (solid curve).}\label{fs8}
	\end{figure}}

\section{Conclusions and final remarks}\label{conclusion}
In this paper, we have considered the cosmological implications of the novel four dimensional Einstein-Gauss-Bonnet gravity. The model is constructed in such a way that the prefactor $d-3$ of the Gauss-Bonnet tensor, where $d$ is the spatial dimensions, is compensated by the Gauss-Bonnet coupling constant. This makes the Gauss-Bonnet tensor to contribute in four dimensions. In this paper, we have considered the cosmology of such a theory in the presence of non-relativistic matter sources. We have used the likelihood analysis to obtain the best values of the model parameter for consistency of the model with observational data.  In $2\sigma$ region of the parameter $\beta$ the late time behavior of the model is equivalent to the standard $\Lambda$CDM model. We have also considered the perturbations of the matter density fluctuations around FRW geometry and obtained growth rate of the matter density perturbations in this model. The 4D EGB theory makes the anisotropic stress parameter to differs from zero. This is in fact a general behavior in higher order modifications of the gravitational action. However, we have shown that the deviations from zero of the anisotropic stress parameter increases as one goes back to early times. This also happens for the effective gravitational constant, larger values of the EGB coupling constant results in a smaller gravitational constant. However for negative values of $\beta$, the effective gravitational constant becomes greater than the standard Newtonian value.

The growth rate of matter density perturbations can be tested by experimental data through the behavior of $f\sigma_8$. Since the growth of gravitational seeds starts in the sub-horizon scale, we have considered sub-horizon limit of the matter dominated epoch in this paper. However the initial condition is modified and the deviation from the $\Lambda$CDM ones is parametrized by the constant $\gamma$.
 We have obtained the best-fit values for all  parameters  $\gamma$, $\sigma_8^0$ and  $\beta$ by maximizing the likelihood function. $1\sigma$ and $2\sigma$ confidence interval for each parameter is presented in Table \ref{T1}. We obtain the value of $\chi_{f\sigma_8}^2/dof$ at the best fit point. With the data sets used in this paper one can easily obtain $\chi_{f\sigma_8}^2/dof$ for $\Lambda$CDM as $0.421$ which has the same order as the value of this parameter in the 4D EGB model.
 
 \subsection*{Acknowledgments}
 We  would like to thank the anonymous referee for very useful comments.

\end{document}